\documentclass[twocolumn]{openjournal}

\usepackage{xcolor}
\usepackage{textgreek}
\usepackage[utf8]{inputenc}
\usepackage[english]{babel}

\usepackage{hyperref}
\hypersetup{
    unicode, 
    colorlinks=true,
    linkcolor=linkcolor,
    citecolor=linkcolor,
    filecolor=linkcolor,
    urlcolor=linkcolor,
}
\usepackage{color,colortbl}
\definecolor{linkcolor}{rgb}{0.0,0.3,0.5}
\usepackage{tensind}
\tensordelimiter{?}
\DeclareGraphicsExtensions{.bmp,.png,.jpg,.pdf}
\usepackage{verbatim}
\usepackage[normalem]{ulem}
\usepackage{orcidlink}
\usepackage{soul}

\begin{document}

\title{Comparing the Architectures of Multiplanet Systems from Kepler, K2, and TESS Data}

\author{Robert Royer III\orcidlink{0000-0002-0900-0192}}
\email{royer@unlv.nevada.edu}
\affiliation{University of Nevada, Las Vegas}
\affiliation{Nevada Center for Astrophysics}

\author{Jason H. Steffen\orcidlink{0000-0003-2202-3847}}
\affiliation{University of Nevada, Las Vegas}
\affiliation{Nevada Center for Astrophysics}

\begin{abstract}

Exoplanet surveys like Kepler, TESS, and K2 have shown that planetary systems are common in our galaxy. These surveys, along with several others, have identified thousands of planetary candidates, with more than five thousand having already been confirmed. Many of these planetary systems host multiple planets.  As we discover more multiplanet systems, notable trends begin to appear in the data.  We use kernel density estimation (KDE) to analyze the period ratios of adjacent planet pairs in multiplanet systems in the most recent Kepler, TESS and K2 data, paying particular attention to pairs in first order mean motion resonance (MMR).  We compare a recent Kepler catalog with the DR25 data release.  We also compare TESS and K2 against this recently released Kepler data.  To verify the significance of our findings against selection bias, we perform Monte Carlo simulations of multiplanet systems in the TESS catalog, finding an excess of planet pairs near the 2 (2:1), and 1.5 (3:2) period ratios, both exceeding the 99\% confidence interval.  We also find a significant peak at the 2.19 period ratio, which exceeds the 90\% confidence interval.  Using a lower limit for period ratios determined by the period of the inner planet proposed in \citet{steffen13}, we identify two planet pairs orbiting M dwarf stars in a very tight ratio.  We also note a likely misidentified planet pair orbiting an FGK type star, which if further study proves to be true, would indicate that only planets orbiting M dwarf stars may violate this limit.
\end{abstract}

\section{INTRODUCTION} \label{sec:intro}

The Kepler Space Telescope \cite{BoruckiKepler} unveiled a wealth of information about the planets residing in our galaxy.  With over five thousand confirmed exoplanets discovered to date, Kepler inspired several follow-up surveys to more fully explore the population of exoplanets; among the most notable of these are the K2 \citep{HowellK2} and TESS (Transiting Exoplanet Survey Satellite) \citep{RickerTESS} missions.  K2 is a continuation of the Kepler mission, extending the lifetime of the satellite after the failure of two of its four reaction wheels prematurely ended its prime mission.  TESS launched in 2018 with the stated goal of observing transiting exoplanets with less than four Earth radii.  As the catalog of exoplanets (and exoplanet candidates) grows we are now able to compare the TESS and K2 data with the previous Kepler findings, gain insights into planetary system architectures, and see if earlier findings from the Kepler mission are seen in these other data.

Studies as early as 2011 have compared the prevalence and properties of multiplanet systems.  \citet{Latham11} analyzed the first four months of Kepler data and showed that approximately 17\% of Kepler planet candidates from that period were found in multiplanet systems.  \citet{WF15} compiled the occurrence rates, inclinations, and orbital spacings of multiplanet systems. Furthermore, \citet{Lissauer2011} showed that most multiplanet observations are of true planets, and not the result of false positive signals.  Several studies also compared early populations of multiplanet system and single planet systems, finding an excess of single planets with orbits less than three days when compared to multiplanet systems (\cite{steffen13}, \cite{SteffenCoughlin16}). They also identify a lower limit for period ratios for planets in short-period orbits \citep{steffen13}.  \citet{Weiss23} investigated the architectures of multiplanet systems, finding that the overall distribution of period ratios is log normal and that smaller exoplanets prefer smaller period ratios. This was corroborated by \citet{Muresan24}, which found that planet pairs smaller than 1 Earth radius tend to be found in period ratios less than 2.

In 2023, \citet{RM23} compared the bulk densities of exoplanet candidates for single and multi systems orbiting M dwarf stars.  They found that planets in a multiplanet system have lower core mass fractions (CMFs) than their counterparts in single planet systems, and that single planet system tend to be less dense than those found in multiplanet systems, implying that there may be a larger number of sub-Neptune planets in single planet systems.  \citet{Liberles23} also find that planets around late K and M dwarfs in multiplanet systems tend to be larger than those in single-planet systems. They find that the larger radii seen in multiplanet systems are not the result of a larger atmosphere from mantle outgassing.  

Studying the catalogs of observed multiplanet systems provides insights into their histories as well.  The capture of planets into resonant chains around their host stars is one of the best models to explain the long term stability of multiplanet systems.  \citet{Huang23} studied planet pairs in mean motion resonance (MMR), where the ratio of the outer planet to the inner planet is a ratio of integers, for insights into the formation history of their stellar systems.  They found that most observed planet pairs can be described by Type I migration (disc migration due to Lindblad resonances).  Their findings also allow them to place an upper limit on the density of the host protoplanetary disc at the time of formation.

Similarly, \citet{Kajtazi23} found that the disc properties have largest impact on the formation of resonant chains.  Their simulations prefer 2:1 and 3:2 resonances \citep{Kajtazi23}.  The 2:1 and 3:2 period ratios were also the focus of the work by \citet{Louden23} studying tidal regimes; finding that the planet bulk density plays a large role in determining the disipation timescale of the system.

A fundamental part of uncovering the mechanisms of planet formation and evolution is understanding the architectures of multiplanet systems, particularly the period ratio distributions of these systems.  Studies such as \citet{Lissauer2011} and \citet{Winn15} found an excess of planet pairs exterior to first order MMRs, and a lack of planet pairs interior to these ratios, particularly the 2:1 period ratio.  In 2015, \citet{Steffen15} identify an excess of planet pairs near the 2.16 period ratio in the early Kepler data.  While \citet{Fang13} find that many multiplanet systems are dynamically full, \citet{Gilbert20} postulate that many lower multiplicity systems host undetected planets.  These findings can inform the search for non-transiting, or other previously unidentified companions, such as the discovery of a small companion interior to Proxima Centauri b at the 2.18 period ratio \citep{Faria2022}. \citet{steffen13} also identified a lower limit for the period ratios of short period planets.  When published the only exception to this rule was the three-planet system of the M-dwarf Kepler-42. A recent study of exoplanet populations by \citet{Howe25} note that Kepler-42 is still an outlier in compact multiplanet systems.

Several studies have also investigated the long-term stability of multiplanet systems.  \citet{Ghosh24} found that the majority of their simulations of high-multiplicity systems undergo periods of instability, resulting in at least one planet lost from the system.  \citet{Volk24} found in their simulations that systems hosting at least one planet pair interior to the 2:1 MMR are more likely to become unstable.  \citet{Deck13} found regions of instability within first-order MMR due to resonance overlap.  Conversely, recent work by \citet{Childs25} compared simulated high multiplicity systems to observations around M-dwarf stars, finding that systems with a period ratio larger than 2 were more likely to have undergone a chaotic event after the dissipation of the protoplanetary disk.  \citet{Chen25} found that such instabilities can limit the number of observed planet pairs near the inner edge of the period ratio distribution. 

Throughout this paper we will examine the Kepler, TESS, and K2 satellites for multiplanet systems and compare the prevalence of systems near first order mean-motion resonances identified in these catalogs.  In Section \ref{sec:data} we describe the three catalogs under inspection. Section \ref{sec:close in} discusses planet pairs that appear below a trend identified in \citet{steffen13}.  We discuss the kernel density estimation (KDE) distributions of these catalogs in Section \ref{sec:dists}, looking at changes in the most up to date Kepler data compared against the DR25 data release, as well as how Kepler compares to TESS and K2.  In Section \ref{sec:stats} we run Kolmgorov-Smirnov and Anderson-Darling tests comparing TESS and K2 to Kepler.  Lastly, in Section \ref{sec:mc} we determine the statistical significance of the peaks seen in the KDEs for TESS multiplanet systems.

\section{DATA} \label{sec:data}
In this study, we analyze the architectures of multiplanet systems in the TESS and K2 exoplanet catalogs and compare these results against published data from the Kepler archive.  We retrieved the catalogs for both TESS and K2 from the NASA Exoplanet Archive on January 18, 2025.  

For the Kepler satellite, we draw our data from \citet{Lissauer24}, which contains 841 multiplanet systems with 1,946 planets, resulting in 1,123 adjacent planet pairs. Due to its large size when compared to either the K2 or TESS catalogs, as well its high number of confirmed exoplanets; we take the Kepler catalog as the standard against which K2 and TESS are compared.

The TESS data used in this study includes observations from sectors 1 through 67, containing 7,358 planetary candidates.  Removing single-planet systems from our study resulted in 220 remaining systems, containing a total of 466 planets, 248 of which are adjacent pairs. Our initial analysis for K2 was performed with the bulk data, in similar fashion to our analysis of Kepler and TESS. However, even calculating the period ratios of K2 systems was complicated by multiple duplicate entries for many planet candidates, and varying amounts of missing parameters throughout the data. For this reason, we use the planets found with the vetting pipeline published by \citet{Zink21}, containing 113 planet candidates in 53 systems, which results in 60 adjacent pairs.  While this pipeline severely limits the number of planetary candidates used for our study when compared to the bulk K2 data, we believe the uniform treatment of the catalog makes it superior to the bulk K2 data, and we show in Section \ref{sec:stats} that this is indeed the case. 

In order to maximize the sample size, we initially discard only confirmed false positive observations, since \citet{Lissauer12} showed that most candidates in multiplanet systems are real planets.  We also restrict our study to exoplanet candidates with less than 20 Earth radii.  Finally, we remove any systems that show an adjacent planet pair having a period ratio of 1.1 or smaller following \cite{Steffen15}.  This cut removes TOI-712 from the TESS catalog, as well as KOIs 284 (Kepler-132), 2248 (Kepler-1913), 6097, and 6242  from the Kepler catalog.  While KOI 284 (Kepler-132) is a system of confirmed exoplanets (albeit orbiting a pair of stars), the adjacent pair of Kepler 132 b/c violates our criteria that no period ratio be less than 1.1 and is therefore removed from our analysis.  Our sample consists of 1,141 adjacent pairs in 823 planetary systems taken from the Kepler catalog, 248 pairs in 220 systems from TESS, and 60 pairs in 53 systems from the K2 catalog.  

\begin{figure}
   \includegraphics[width=\columnwidth]{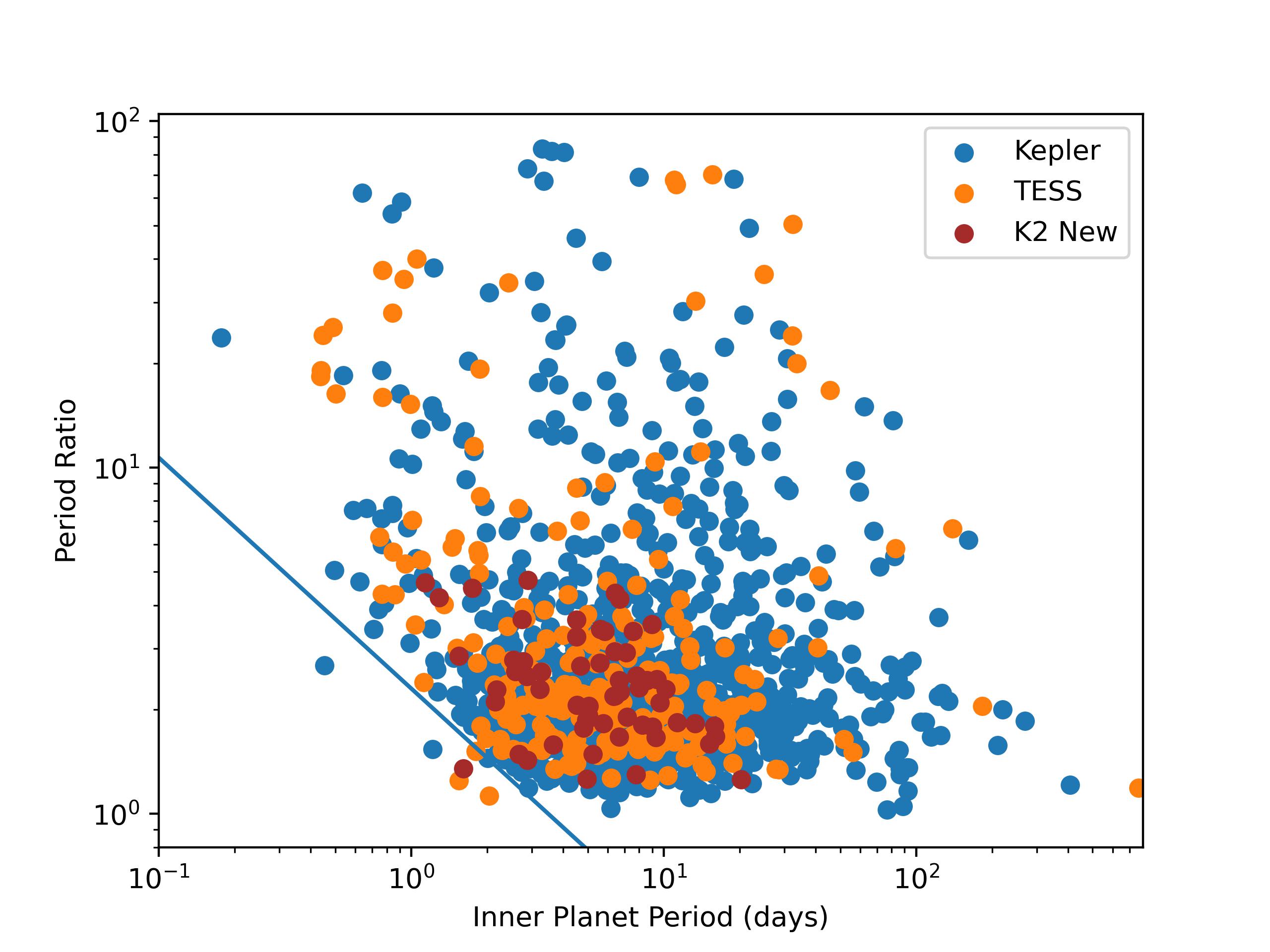}
    \caption{Scatter plot showing the period ratio as a function of the inner planet period, in days, for Kepler (blue), TESS (green), and K2 (red).  The solid blue line in the lower left corner indicates a lower limit to observed period ratios (\ref{eq:1}) as discussed in Section \ref{sec:close in}.}
    \label{catalogs}
\end{figure}

\section{CLOSE IN PLANETS} \label{sec:close in}
One of the most interesting revelations to arise from earlier studies of the Kepler catalog is the presence of a lower limit to the observed period ratios of multiplanet systems.  First identified in \citet{steffen13}, this lower limit is characterized by the equation
\begin{equation} \label{eq:1}
   P = 2.3 P_1^{-2/3} 
\end{equation}
where $P_1$ is the period of the inner planet in days and $P$ is the period ratio of the two planets (a ratio always larger than unity).  Although no prescription of a lower limit is made, this can also be seen in the recent work by \citet{Howe25}. 

Earlier studies \citep{steffen13} identified only two exceptions to this rule, both occurring in the same system, the adjacent planet pairs in the three-planet system of Kepler 42.  This was notable because Kepler 42 is one of the few multiplanet M dwarfs to be observed by the Kepler mission and, actually, one of the smallest stars in the catalog (0.17 Solar radii, 0.14 Solar masses).

The most updated catalogs from Kepler show only one new planet pair (KOI 2248.01/.04) that appears to violate this rule, but we call this observation into question.  A recent study by \citet{Lissauer24} shows that this planet pair are likely to be observations of planets orbiting two different stars with very little angular separation.   Due to the questionable nature of KOI 2248 we have removed it from consideration in our analysis.

When we look at the TESS data, we see two planet pairs with period ratio values smaller than what is predicted by equation \ref{eq:1}. TOI 6022.01/.02 has a period ratio of 1.25 and an inner orbital period of 1.55 days, which is much lower than the predicted bound on that ratio of 1.72.  TOI 797.03/.01 has a period ratio of 1.52 and an inner orbital period of 1.80 days.  This period ratio is very close to the period ratio of 1.55 predicted by Equation \ref{eq:1}.

TOI 6022 has two identified planet candidates while TOI 797 hosts three planet candidates (of which the innermost pair violates equation 2).  It is unsurprising that we find violations to Equation 2 orbiting TOIs 797 (0.47 Solar radii) and 6022 (0.36 Solar radii), as they orbit M dwarf stars, much like Kepler 42. It is worth noting that only one planet pair orbiting TOI 797 violates this rule, whereas both of the planet pairs in the Kepler 42 system violate this lower limit.

\section{ADJACENT PLANET PAIR DISTRIBUTIONS} \label{sec:dists}
In order to identify the period ratio distributions of adjacent planet pairs in the three catalogs we are investigating (Kepler, TESS, and K2) we first identify all surveyed star systems with more than one observed planet candidate. Then, after checking that none of these candidates are marked with a false positive disposition, we construct a list of star systems with the orbital periods of the planet candidates sorted from smallest to largest.  We next take these orbital periods and calculate the period ratios of adjacent planet pairs, using the definition 

\begin{equation}
P = P_{n+1}/P_n.
\end{equation}

\begin{figure}
        \centering
        \includegraphics[width=\columnwidth]{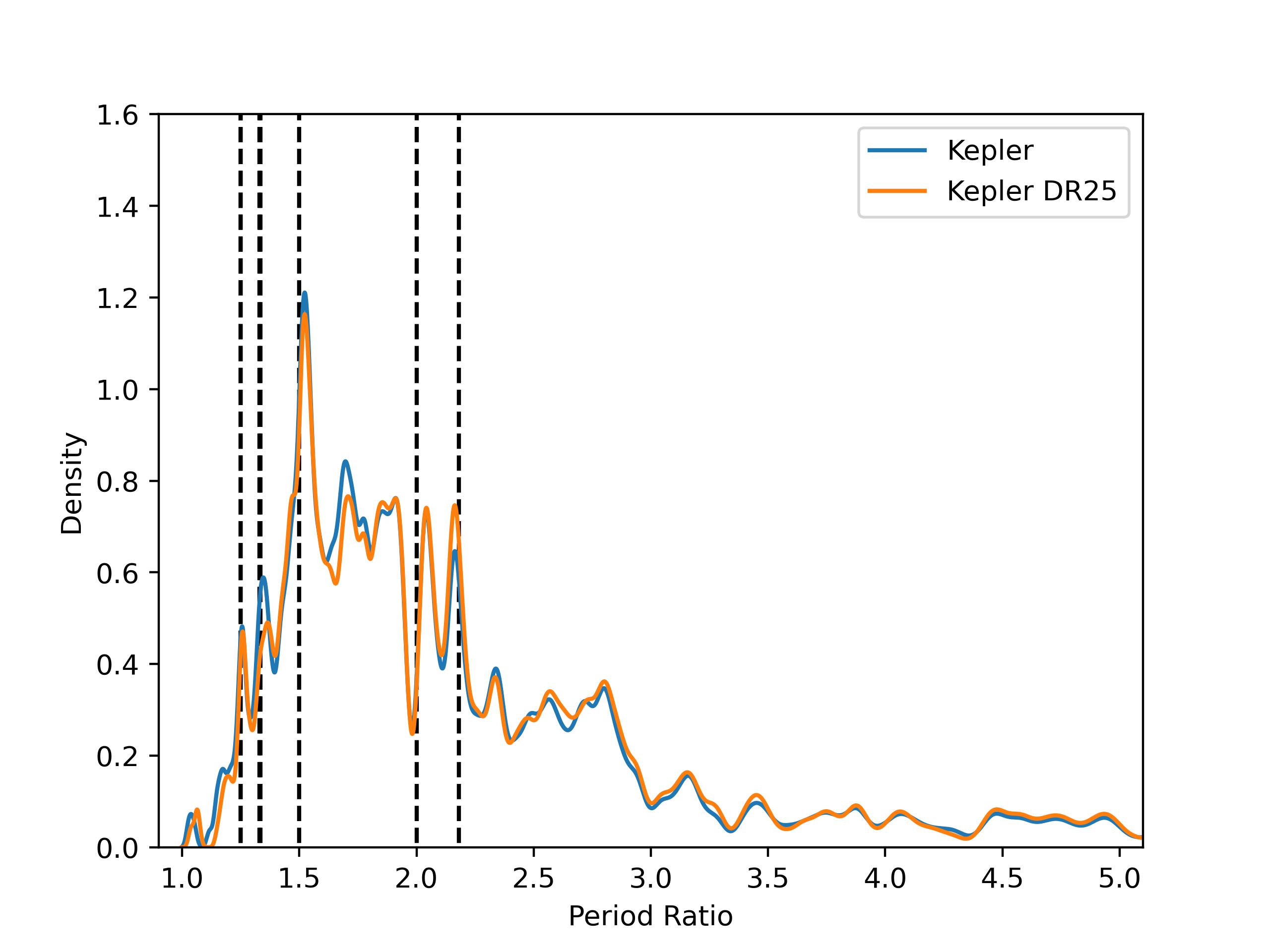}
        \label{fig:kep KDE}
    \hfill
        \centering
        \includegraphics[width=\columnwidth]{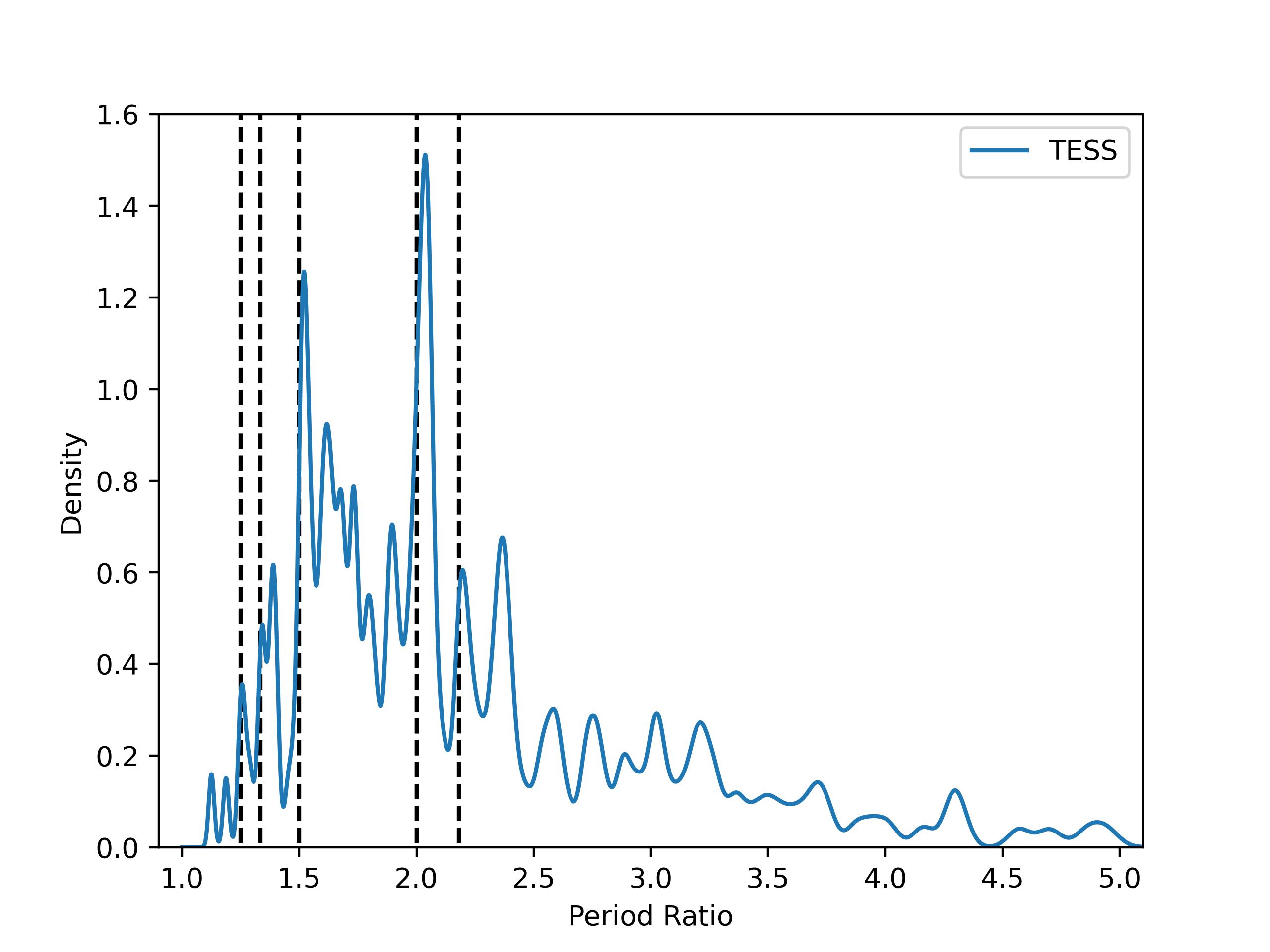}
        \label{fig:TESS KDE}
    \hfill
        \centering
        \includegraphics[width=\columnwidth]{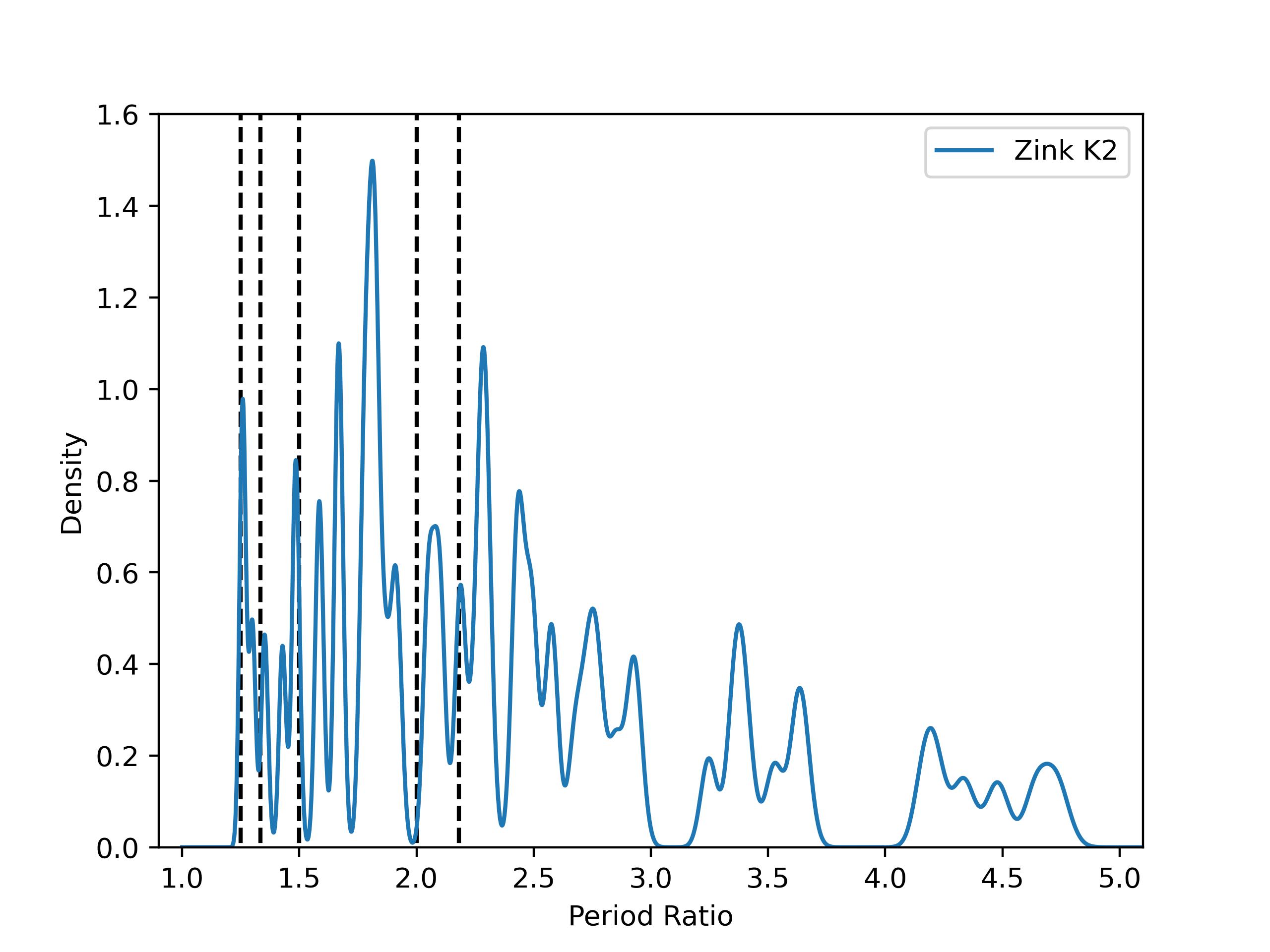}
        \label{fig:K2 KDE}

    \caption{Kernel Density Estimation using the normal distribution of Kepler (top), TESS (middle), and K2 (bottom) for period ratios of adjacent planet pairs.  The vertical dashed lines denote the locations of the 5/4, 4/3, 3/2, 2/1, and 2.19 period ratios. Top figure compares adjacent planet pairs between the Lissauer and DR25 catalogs.}
    \label{fig:kdes}
\end{figure}

To quantify the distributions of each catalog, we visualize the period ratios in a kernel density estimation (KDE) plot (\ref{fig:kdes}), using the normal distribution as the kernel, with a variable bandwidth equal to 0.015 times the period ratio value. We choose a variable bandwidth for our kernel so that we can more precisely identify small features in the distribution at low period ratio values, while simultaneously reducing some of the noise present at large period ratios, similar to the method in \citet{Steffen15}.  To provide for more direct comparison between the catalogs, we normalize the scaling so that each KDE under inspection integrates to one.  

Using the most recent Kepler catalog, published by \citet{Lissauer24} (hereafter the Lissauer catalog) we see a similar distribution of planet ratios as compared to previous data releases, such as the DR25 data release.  The most prominent feature of the distribution is a large spike located at the 3:2 MMR, which aligns with the maximum peak identified in \citet{Steffen15}.  Comparing this peak to the DR25 data, we see a slight increase in density.  Other interesting features of the distribution we see in both the older and more recent Kepler catalogs are located at the 2:1, 4:3, and 5:4 MMRs, and the 2.16 period ratio.

At the 2:1 MMR we see a lack of planet pairs just interior to this ratio, as well as a peak just exterior to it.  The density of this feature is largely unchanged since the DR25 release, as is the density of the peak at the 5:4 MMR.  The peak that we see at the 4:3 MMR has increased in density with the Lissauer data.  The period ratio of 2.16 is quite interesting, as it is one of the few features that has a lower density when comparing the Lissauer data to the DR25 release.

The distribution of period ratios in the TESS catalog shows a large number of planet pairs just outside the 2:1 MMR ratio, as shown in the KDE plot \ref{fig:kdes}.  This period ratio shows a much higher occurrence rate than is seen in the Kepler catalog, almost twice as high when comparing their respective KDE's.  We see another large peak at the 3:2 MMR, which corresponds to the largest peak we see in the Kepler catalog.  The density of planet pairs at this peak is slightly lower than Kepler, but remains significantly higher than the rest of the TESS distribution. We see a shift in the peak near 2.2 when compared to the Kepler data.  While the peak location in the Kepler data is 2.16, this peak appears at 2.19 in the TESS data.  The fact that this peak is present in both catalogs reinforces that it is astrophysical in nature rather than statistical noise.

The K2 distribution differs from Kepler and TESS in the locations of its most prominent peaks.  While both Kepler and TESS show a high density for planet pairs at 1.5, this period ratio show much lower densities in our KDE for K2.  In the case of the 2.0 ratio, the density is very near to zero.  Unlike the Kepler and TESS catalogs, we see no large excess of planet pairs exterior to 2.0, though this may be due to the small sample size of K2. We also see a small peak at the 2.19 period ratio, but it is much less prominent than we see with Kepler and TESS.  The distribution of planet pairs for the K2 catalog also  shows a large number of pairs with 1.25, 1.75, and 2.25 period ratios. The peak just exterior to 1.75 shows the highest density in the K2 data.  This is not seen in the Kepler data, which shows a very slight dip in the population density for this ratio.  Similarly, the peak at 2.25 in the K2 data shows a small trough in the Kepler data.
\section{STATISTICAL ANALYSIS OF BULK K2 DATA} \label{sec:stats}
To determine the likelihood that these catalogs represent samples taken from the same population, we ran Kolmogorov-Smirnov (KS) and Anderson-Darling (AD) tests between the distribution of orbital periods in the Kepler catalog and the TESS and K2 catalogs.  For each test, we chose a maximum period value ($P_{max}$), and removed any planetary candidates with orbital periods greater than the cutoff value.  We then iterated the maximum period from 50 to 1 days in 0.1 day increments, and recorded the p-value at each increment.  We do not test the agreement of planets with orbital periods greater than 50 days, as this is more than twice the duration of TESS sector observations, increasing the chance for period aliasing. 

\begin{figure}
    \centering
        \centering
        \includegraphics[width=\columnwidth]{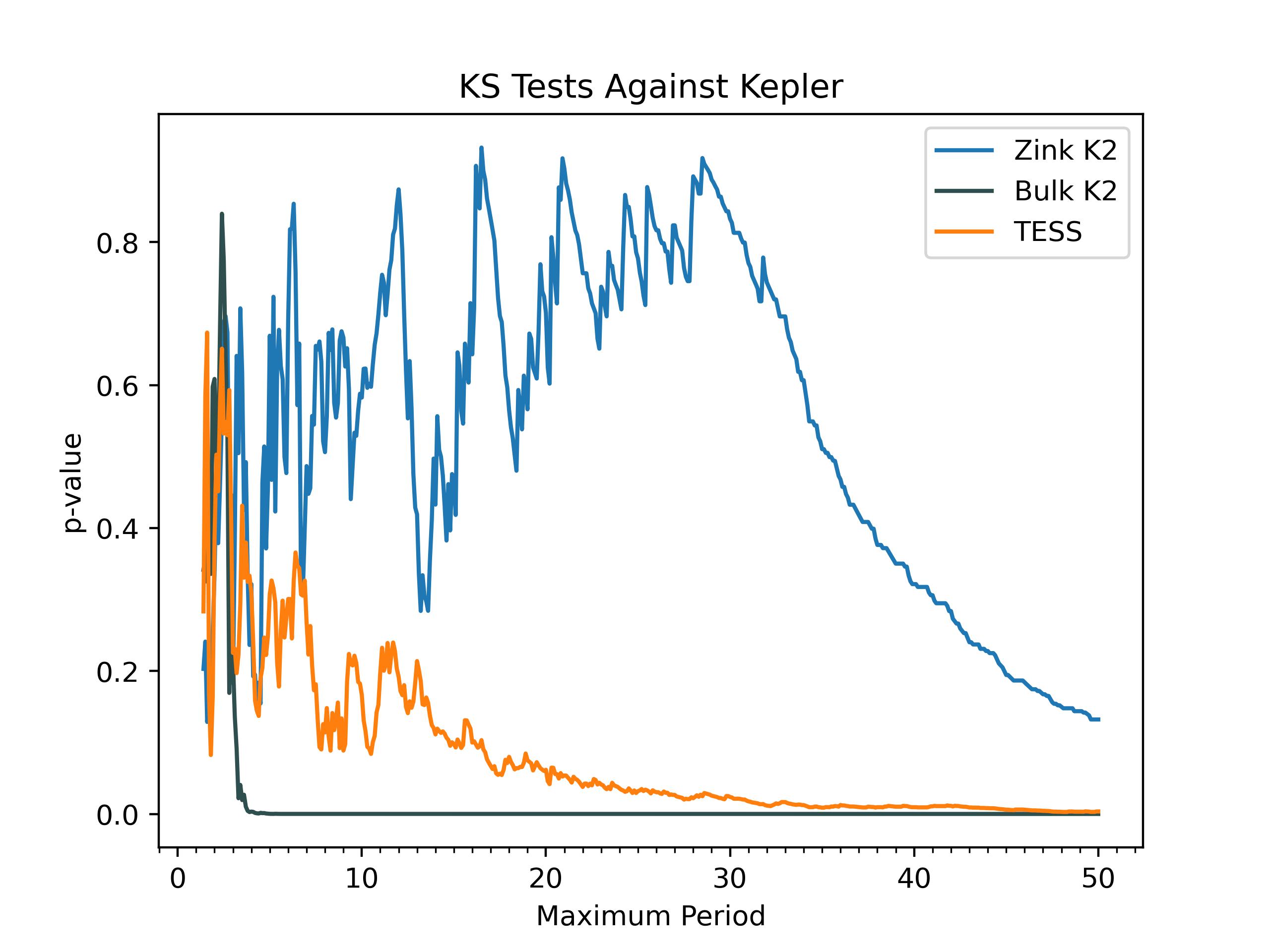}
        \label{fig:KS}
    \hfill
        \centering
        \includegraphics[width=\columnwidth]{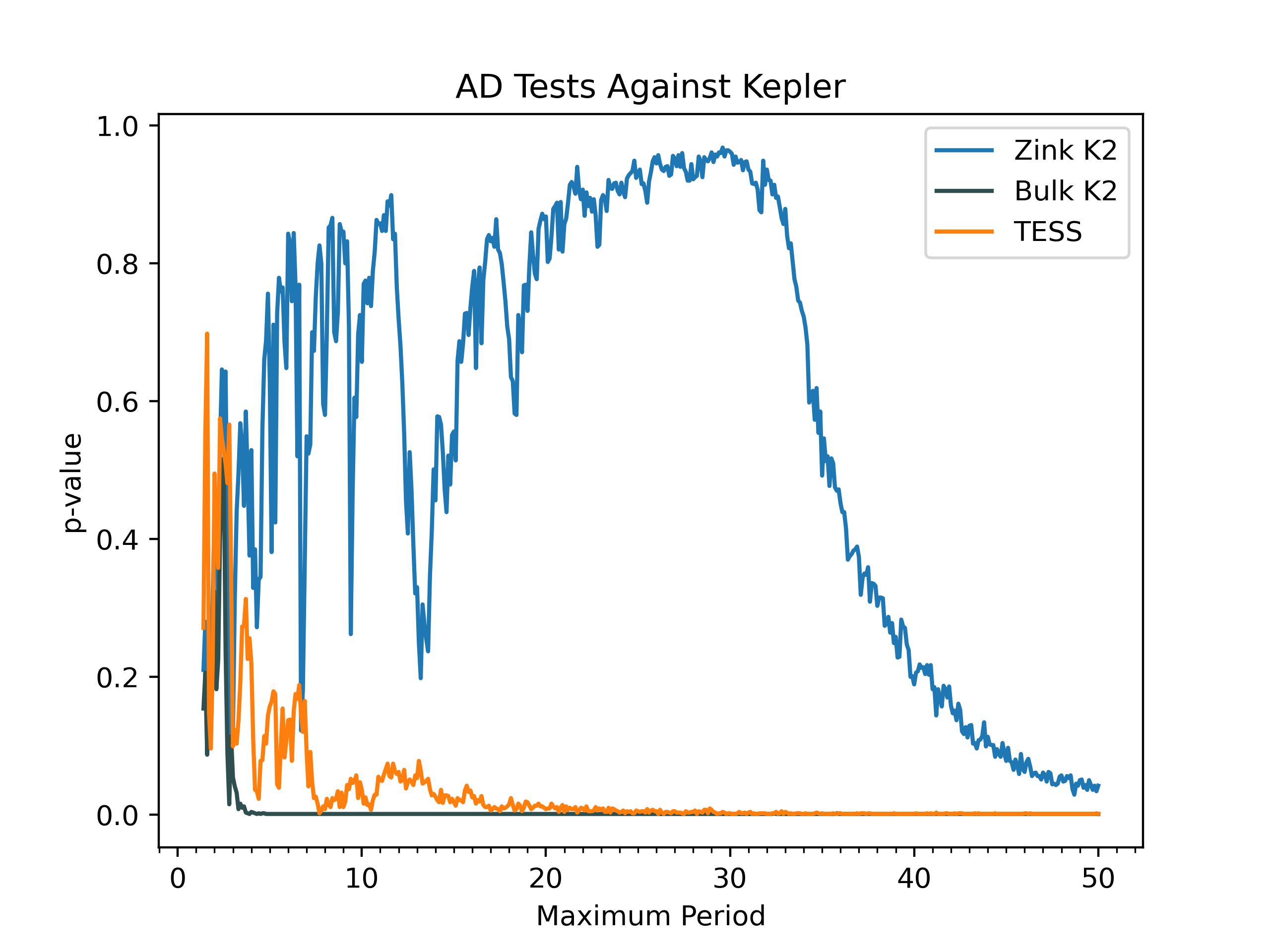}
        \label{fig:AD}
    \caption{P-value results for Kolmgorov-Smirnov (a) and Anderson-Darling (b) tests comparing the orbital periods of the TESS and K2 catalogs against the Kepler catalog. For each maximum period value, planets with larger orbital periods are removed from consideration.}
    \label{fig:stats}
\end{figure}

There is moderate agreement when comparing the Kepler and TESS catalogs, although the agreement is noticeably lower for the AD tests, as seen in Figure \ref{fig:stats}.  The p-values for both the KS and AD tests are shown in the upper and lower panels, respectively.  For both the KS and AD cases, p-values reach a maximum value for $P_{max}$ less than 3 days, implying that the samples in this period range have the greatest probability of being from the same underlying population of planets.  This may, however, be a result of the small number of observed candidates in this range.  A more meaningful peak appears in the KS test at 13 day periods, with p-values diminishing as longer orbital periods are included in the analysis. However, we expect this level of consistency due to the majority of orbital period values in both catalogs falling below 10 days, and in the case of TESS, with very few planetary candidates observed after 30 days.  The moderate p-values that we see in the range of 3 to 30 days do not allow us to draw any meaningful conclusions about the underlying populations of planets for these catalogs.  

The data from the Kepler and bulk K2 catalogs show very little agreement in the statistical properties of their sample populations for both the KS and AD tests.  The majority of maximum period values produce p-values for both tests return the lower floor value of 0.001 for each test.  

Both the KS and AD tests of Kepler and K2 show notably higher p-values for maximum periods between one and five days suggesting that K2 is more capable of detecting planets in this small range of periods than on longer orbital periods. There is a very noticeable maximum for both tests at $P_{max}$ equal to two days.  But this is, again, most likely due to the very small sample size in the K2 data at these low period values.  

The Kepler catalog has identified over six times as many multiplanet systems as K2.  This disparity in catalog size may be the cause of the lack of agreement between the two catalogs due to the small size of the K2 sample. To test this hypothesis we randomly selected 126 systems from the Kepler catalog, and this subset was compared against the remaining star systems in the Kepler multiplanet catalog for differing $P_{max}$ values.  Note that we opted to constrain our sub-samples via the number of star systems, rather than the number of planets, in the catalog due to the fact that single planet systems have already been removed.  After randomly selecting our sub-sample of Kepler systems, we then compared it to the remaining  Kepler population using KS and AD tests, again, decreasing $P_{max}$ from 50 to 1 days.  This process was repeated for 1,000 iterations to determine both an average p-value for the random sub-samples and to look for any trends in the simulated data.  

The results of our random sampling show a wide range in p-values for both the KS and AD tests.  In both cases, the average value is much higher than is seen when comparing the Kepler and bulk K2 catalogs, suggesting that it is unlikely that the low statistical agreement seen when comparing these catalogs are not the result of a disparity in sample size.  Comparing the KS tests for a maximum of 25-day periods, all of our 1,000 random samples had p-values greater than we see between the Kepler and K2 catalogs ($1.9\times10^{-4}$).  We note that a small number of samples (4 out of 1,000 samples) do show p-values on the order of those seen between Kepler and K2.  

As an alternative to the bulk K2 catalog found on the Exoplanet Archive, we examine the \citet{Zink21} catalog.  This catalog, released in 2021, aims to provide a level of vetting and completeness that is not seen in the bulk K2 data.  It also removed the need for visual inspection of the light curve data, reducing subjectivity in the verification of transit signals.  These more robust vetting techniques come with a severe trade-off in terms of sample size, reducing the number of multiplanet systems in the catalog from 126 in the bulk data to 53 systems in the updated catalog.  This also reduces the number of adjacent planet pairs from 182 to 60 pairs.

Statistical comparisons between the Kepler and Zink K2 catalogs show a higher level of agreement than is seen with the original catalog.  P-values returned for the Anderson-Darling tests between the Zink K2 catalog and Kepler are above 0.5 for most $P_{max}$ values and reach over 0.9 for $P_{max}$ values between 25 and 30 days. This is in striking comparison to the values of 0.0001 returned for the same $P_{max}$ values in the bulk K2 catalog.  Similarly, the results of the Kolmogorov-Smirnov tests show more agreement for the reduced Zink catalog than the bulk catalog. The returned p-values increase as a function of $P_{max}$ until reaching a maximum value of 0.95 for $P_{max}$ equal to 30 days, although we expect this given that the largest period value in the K2 catalog is approximately 31 days.

One important take-away point of this analysis is the care that should be taken if the bulk K2 catalog is used for planet demographic statistics.  Assuming that the Kepler catalog is our best standard for transiting planet demographics, then the substantial disagreement between the Kepler catalog and the K2 catalog indicates that there may be important gaps in the K2 planet candidate populations stemming from the different approaches to the observation strategy, analyzing the data, and producing the catalog. The improved agreement with the Zink catalog would motivate its use for such studies over the general K2 catalog.

\section{DETERMINING PEAK SIGNIFICANCE} \label{sec:mc}
To quantify the significance of the peaks observed in our KDE distributions, we perform Monte Carlo (MC) simulations for a selection of planet pairs from an assumed underlying distribution.  To achieve this, we use the KDE values for each catalog, taken with a large enough bandwidth that the distributions become unimodal, and randomly sample period ratio values to build a sample set that is the same length as the catalog of interest.  To determine the peak and valley locations in each realization we then analytically fit each set of three adjacent data points with a quadratic function to determine local minima and maxima.  After completing this process for a total of one thousand MC sample sets, we use the appropriate values from this database to build our confidence intervals (the 5 and 95 percentile values to make the 90\% confidence interval, and the 0.5 and 99.5 percentile values to make the 99\% confidence interval). 

When looking at the confidence intervals for the TESS data, we see several peaks and troughs that exceed even the 99\% confidence interval.  The most striking of these are the peaks located just exterior to the 3:2 (1.5) and 2:1 MMRs. Both have density values approaching twice that of the 99\% confidence interval, although both of these peaks have been observed to be significant in prior studies of the Kepler catalog \citep{Steffen15}.  Another peak that has been previously identified to be significant is the peak at the 2.19 period ratio.  While the corresponding peak (2.16) exceeds the 99\% confidence interval in the Kepler data, it rises to only the 90\% confidence interval in the TESS data.

\begin{figure}
    \centering
    \includegraphics[width=\columnwidth]{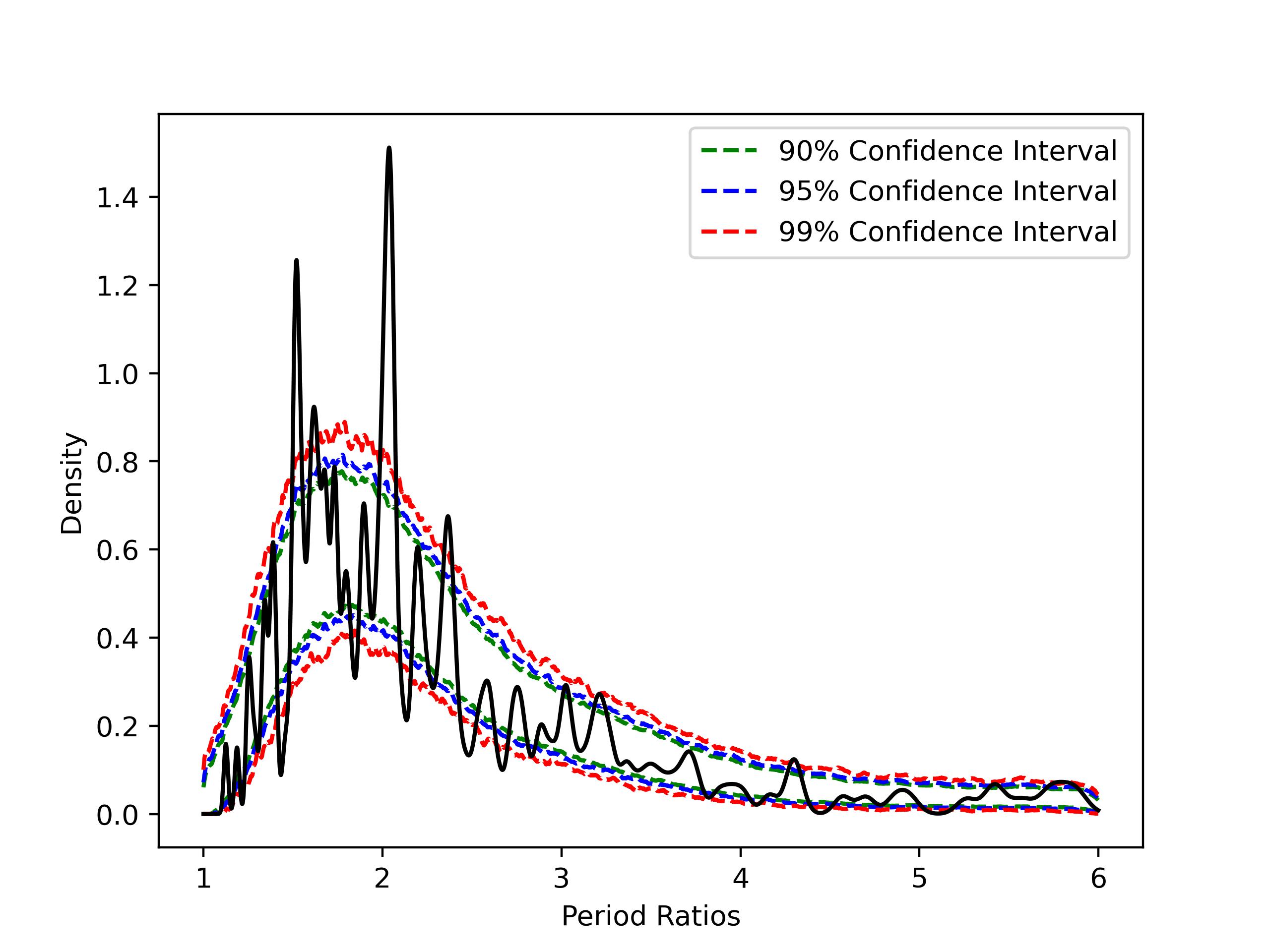}
    \label{fig:TESS mc}
    \caption{Significance estimates for the peaks of the KDE of the period ratio distribution for TESS multiplanet systems using 1000 realizations of the nominal (unimodal) distribution.  The 90th (green), 95th (blue), and 99th (red) intervals are indicated.}
\end{figure}

\section{CONCLUSION}
The number of planets in multiplanet systems that Kepler has identified remains far larger than either the TESS or K2 catalogs; although TESS continues to make strides in this respect.  TESS has observed two planet pairs that violate Equation \ref{eq:1}, which describes the minimum period ratio for planet pairs as a function of the period of the inner planet that appeared to be respected by planetary systems in the Kepler data. Both of these pairs reside around M dwarf stars. The only new addition in the Kepler catalog that violates this trend, KOI-2248 (inner period of 2.64 days and a period ratio of 1.06) is likely a blended signal from a close binary system. If further studies show this to be true, it will further reinforce the presence of this lower limit.

We see no major changes to the period ratio distribution of the Kepler catalog when comparing between DR25 and the catalog released by \citet{Lissauer22}. The 3:2 MMR peak remains the dominant feature of the distribution, as well as a slightly lower density at the 2:1 MMR than is seen in previous studies of the Kepler catalog.  Much like Kepler, TESS shows an abundance of planet pairs at the 3:2 MMR; but the most striking feature is the large excess of planets at the 2:1 period ratio.  

When we compare the sample populations using statistical methods, such as AD and KS tests, we see moderate agreement between the TESS and Kepler catalogs, but very little agreement between Kepler and the bulk K2 data. The K2 catalog generated by the vetting pipeline by \citet{Zink21} greatly increases the agreement between the Kepler and K2 catalogs, but also removes most K2 candidates.  

When looking at the statistical significance of the peaks via Monte Carlo simulations we see that both the 3:2 and 2:1 MMR peaks are significant in the TESS data above the 99\% confidence interval. As seen in the Kepler data, the peak at 2.19 is also of weak significance in the TESS data, exceeding the 90\% confidence interval.

\begin{acknowledgments}
We thank the members of the Kepler, K2, and TESS mission teams for their dedication to finding and vetting the many transit signals that have made this work possible.
\end{acknowledgments}

\bibliographystyle{aasjournal}
\bibliography{Multiplanet_Architectures}{}

\begin{thebibliography}{}
\expandafter\ifx\csname natexlab\endcsname\relax\def\natexlab#1{#1}\fi
\providecommand{\url}[1]{\href{#1}{#1}}
\providecommand{\dodoi}[1]{doi:~\href{http://doi.org/#1}{\nolinkurl{#1}}}
\providecommand{\doeprint}[1]{\href{http://ascl.net/#1}{\nolinkurl{http://ascl.net/#1}}}
\providecommand{\doarXiv}[1]{\href{https://arxiv.org/abs/#1}{\nolinkurl{https://arxiv.org/abs/#1}}}

\bibitem[{Borucki {et~al.}(2010)Borucki, Koch, Basri, Batalha, Brown, Caldwell, Caldwell, Christensen-Dalsgaard, Cochran, DeVore, Dunham, Dupree, Gautier, Geary, Gilliland, Gould, Howell, Jenkins, Kondo, Latham, Marcy, Meibom, Kjeldsen, Lissauer, Monet, Morrison, Sasselov, Tarter, Boss, Brownlee, Owen, Buzasi, Charbonneau, Doyle, Fortney, Ford, Holman, Seager, Steffen, Welsh, Rowe, Anderson, Buchhave, Ciardi, Walkowicz, Sherry, Horch, Isaacson, Everett, Fischer, Torres, Johnson, Endl, MacQueen, Bryson, Dotson, Haas, Kolodziejczak, Cleve, Chandrasekaran, Twicken, Quintana, Clarke, Allen, Li, Wu, Tenenbaum, Verner, Bruhweiler, Barnes, \& Prsa}]{BoruckiKepler}
Borucki, W.~J., Koch, D., Basri, G., {et~al.} 2010, Science, 327, 977, \dodoi{10.1126/science.1185402}

\bibitem[{{Chen} {et~al.}(2025){Chen}, {Cardenas}, {Bonifacio}, {Hall}, {Kang}, \& {Tamayo}}]{Chen25}
{Chen}, K., {Cardenas}, O., {Bonifacio}, B., {et~al.} 2025, \apj, 982, 100, \dodoi{10.3847/1538-4357/adae8a}

\bibitem[{{Childs} {et~al.}(2025){Childs}, {Hua}, {Martin}, {Yang}, \& {Geller}}]{Childs25}
{Childs}, A.~C., {Hua}, A. P.~S., {Martin}, R.~G., {Yang}, C.-C., \& {Geller}, A.~M. 2025, \apj, 982, 111, \dodoi{10.3847/1538-4357/adbb53}

\bibitem[{{Deck} {et~al.}(2013){Deck}, {Payne}, \& {Holman}}]{Deck13}
{Deck}, K.~M., {Payne}, M., \& {Holman}, M.~J. 2013, \apj, 774, 129, \dodoi{10.1088/0004-637X/774/2/129}

\bibitem[{{Fang} \& {Margot}(2013)}]{Fang13}
{Fang}, J., \& {Margot}, J.-L. 2013, \apj, 767, 115, \dodoi{10.1088/0004-637X/767/2/115}

\bibitem[{{Faria} {et~al.}(2022){Faria}, {Su{\'a}rez Mascare{\~n}o}, {Figueira}, {Silva}, {Damasso}, {Demangeon}, {Pepe}, {Santos}, {Rebolo}, {Cristiani}, {Adibekyan}, {Alibert}, {Allart}, {Barros}, {Cabral}, {D'Odorico}, {Di Marcantonio}, {Dumusque}, {Ehrenreich}, {Gonz{\'a}lez Hern{\'a}ndez}, {Hara}, {Lillo-Box}, {Lo Curto}, {Lovis}, {Martins}, {M{\'e}gevand}, {Mehner}, {Micela}, {Molaro}, {Nunes}, {Pall{\'e}}, {Poretti}, {Sousa}, {Sozzetti}, {Tabernero}, {Udry}, \& {Zapatero Osorio}}]{Faria2022}
{Faria}, J.~P., {Su{\'a}rez Mascare{\~n}o}, A., {Figueira}, P., {et~al.} 2022, \aap, 658, A115, \dodoi{10.1051/0004-6361/202142337}

\bibitem[{{Ghosh} \& {Chatterjee}(2024)}]{Ghosh24}
{Ghosh}, T., \& {Chatterjee}, S. 2024, \mnras, 527, 79, \dodoi{10.1093/mnras/stad2962}

\bibitem[{{Gilbert} \& {Fabrycky}(2020)}]{Gilbert20}
{Gilbert}, G.~J., \& {Fabrycky}, D.~C. 2020, \aj, 159, 281, \dodoi{10.3847/1538-3881/ab8e3c}

\bibitem[{{Howe} {et~al.}(2025){Howe}, {Becker}, {Stark}, \& {Adams}}]{Howe25}
{Howe}, A.~R., {Becker}, J.~C., {Stark}, C.~C., \& {Adams}, F.~C. 2025, \aj, 169, 149, \dodoi{10.3847/1538-3881/adabdb}

\bibitem[{{Howell} {et~al.}(2014){Howell}, {Sobeck}, {Haas}, {Still}, {Barclay}, {Mullally}, {Troeltzsch}, {Aigrain}, {Bryson}, {Caldwell}, {Chaplin}, {Cochran}, {Huber}, {Marcy}, {Miglio}, {Najita}, {Smith}, {Twicken}, \& {Fortney}}]{HowellK2}
{Howell}, S.~B., {Sobeck}, C., {Haas}, M., {et~al.} 2014, \pasp, 126, 398, \dodoi{10.1086/676406}

\bibitem[{{Huang} \& {Ormel}(2023)}]{Huang23}
{Huang}, S., \& {Ormel}, C.~W. 2023, \mnras, 522, 828, \dodoi{10.1093/mnras/stad1032}

\bibitem[{{Kajtazi} {et~al.}(2023){Kajtazi}, {Petit}, \& {Johansen}}]{Kajtazi23}
{Kajtazi}, K., {Petit}, A.~C., \& {Johansen}, A. 2023, \aap, 669, A44, \dodoi{10.1051/0004-6361/202244460}

\bibitem[{{Latham} {et~al.}(2011){Latham}, {Rowe}, {Quinn}, {Batalha}, {Borucki}, {Brown}, {Bryson}, {Buchhave}, {Caldwell}, {Carter}, {Christiansen}, {Ciardi}, {Cochran}, {Dunham}, {Fabrycky}, {Ford}, {Gautier}, {Gilliland}, {Holman}, {Howell}, {Ibrahim}, {Isaacson}, {Jenkins}, {Koch}, {Lissauer}, {Marcy}, {Quintana}, {Ragozzine}, {Sasselov}, {Shporer}, {Steffen}, {Welsh}, \& {Wohler}}]{Latham11}
{Latham}, D.~W., {Rowe}, J.~F., {Quinn}, S.~N., {et~al.} 2011, \apjl, 732, L24, \dodoi{10.1088/2041-8205/732/2/L24}

\bibitem[{{Liberles} {et~al.}(2023){Liberles}, {Dittmann}, {Elardo}, \& {Ballard}}]{Liberles23}
{Liberles}, B.~T., {Dittmann}, J.~A., {Elardo}, S.~M., \& {Ballard}, S. 2023, arXiv e-prints, arXiv:2312.05809, \dodoi{10.48550/arXiv.2312.05809}

\bibitem[{{Lissauer} \& {Kepler Science Team}(2012)}]{Lissauer12}
{Lissauer}, J.~J., \& {Kepler Science Team}. 2012, in AAS/Division for Planetary Sciences Meeting Abstracts, Vol.~44, AAS/Division for Planetary Sciences Meeting Abstracts \#44, 100.02

\bibitem[{{Lissauer} {et~al.}(2022){Lissauer}, {Rowe}, {Jontof-Hutter}, \& {Fabrycky}}]{Lissauer22}
{Lissauer}, J.~J., {Rowe}, J.~F., {Jontof-Hutter}, D., \& {Fabrycky}, D.~C. 2022, in LPI Contributions, Vol. 2687, Exoplanets in Our Backyard 2, 3027

\bibitem[{{Lissauer} {et~al.}(2024){Lissauer}, {Rowe}, {Jontof-Hutter}, {Fabrycky}, {Ford}, {Ragozzine}, {Steffen}, \& {Nizam}}]{Lissauer24}
{Lissauer}, J.~J., {Rowe}, J.~F., {Jontof-Hutter}, D., {et~al.} 2024, PSJ, 5, 152, \dodoi{10.3847/PSJ/ad0e6e}

\bibitem[{Lissauer {et~al.}(2011)Lissauer, Ragozzine, Fabrycky, Steffen, Ford, Jenkins, Shporer, Holman, Rowe, Quintana, Batalha, Borucki, Bryson, Caldwell, Carter, Ciardi, Dunham, Fortney, Gautier, Howell, Koch, Latham, Marcy, Morehead, \& Sasselov}]{Lissauer2011}
Lissauer, J.~J., Ragozzine, D., Fabrycky, D.~C., {et~al.} 2011, The Astrophysical Journal Supplement Series, 197, 8, \dodoi{10.1088/0067-0049/197/1/8}

\bibitem[{{Louden} {et~al.}(2023){Louden}, {Laughlin}, \& {Millholland}}]{Louden23}
{Louden}, E.~M., {Laughlin}, G.~P., \& {Millholland}, S.~C. 2023, \apjl, 958, L21, \dodoi{10.3847/2041-8213/ad0843}

\bibitem[{{Muresan} {et~al.}(2024){Muresan}, {Persson}, \& {Fridlund}}]{Muresan24}
{Muresan}, A., {Persson}, C.~M., \& {Fridlund}, M. 2024, \aap, 692, A122, \dodoi{10.1051/0004-6361/202451353}

\bibitem[{{Ricker} {et~al.}(2015){Ricker}, {Winn}, {Vanderspek}, {Latham}, {Bakos}, {Bean}, {Berta-Thompson}, {Brown}, {Buchhave}, {Butler}, {Butler}, {Chaplin}, {Charbonneau}, {Christensen-Dalsgaard}, {Clampin}, {Deming}, {Doty}, {De Lee}, {Dressing}, {Dunham}, {Endl}, {Fressin}, {Ge}, {Henning}, {Holman}, {Howard}, {Ida}, {Jenkins}, {Jernigan}, {Johnson}, {Kaltenegger}, {Kawai}, {Kjeldsen}, {Laughlin}, {Levine}, {Lin}, {Lissauer}, {MacQueen}, {Marcy}, {McCullough}, {Morton}, {Narita}, {Paegert}, {Palle}, {Pepe}, {Pepper}, {Quirrenbach}, {Rinehart}, {Sasselov}, {Sato}, {Seager}, {Sozzetti}, {Stassun}, {Sullivan}, {Szentgyorgyi}, {Torres}, {Udry}, \& {Villasenor}}]{RickerTESS}
{Ricker}, G.~R., {Winn}, J.~N., {Vanderspek}, R., {et~al.} 2015, Journal of Astronomical Telescopes, Instruments, and Systems, 1, 014003, \dodoi{10.1117/1.JATIS.1.1.014003}

\bibitem[{{Rodr{\'\i}guez Mart{\'\i}nez} {et~al.}(2023){Rodr{\'\i}guez Mart{\'\i}nez}, {Martin}, {Gaudi}, {Schulze}, {Asnodkar}, {Boley}, \& {Ballard}}]{RM23}
{Rodr{\'\i}guez Mart{\'\i}nez}, R., {Martin}, D.~V., {Gaudi}, B.~S., {et~al.} 2023, \aj, 166, 137, \dodoi{10.3847/1538-3881/aced9a}

\bibitem[{{Steffen} \& {Coughlin}(2016)}]{SteffenCoughlin16}
{Steffen}, J.~H., \& {Coughlin}, J.~L. 2016, Proceedings of the National Academy of Science, 113, 12023, \dodoi{10.1073/pnas.1606658113}

\bibitem[{Steffen \& Farr(2013)}]{steffen13}
Steffen, J.~H., \& Farr, W.~M. 2013, The Astrophysical Journal, 774, L12, \dodoi{10.1088/2041-8205/774/1/l12}

\bibitem[{{Steffen} \& {Hwang}(2015)}]{Steffen15}
{Steffen}, J.~H., \& {Hwang}, J.~A. 2015, mnras, 448, 1956, \dodoi{10.1093/mnras/stv104}

\bibitem[{{Volk} \& {Malhotra}(2024)}]{Volk24}
{Volk}, K., \& {Malhotra}, R. 2024, \aj, 167, 271, \dodoi{10.3847/1538-3881/ad3de5}

\bibitem[{{Weiss} {et~al.}(2023){Weiss}, {Millholland}, {Petigura}, {Adams}, {Batygin}, {Block}, \& {Mordasini}}]{Weiss23}
{Weiss}, L.~M., {Millholland}, S.~C., {Petigura}, E.~A., {et~al.} 2023, in Astronomical Society of the Pacific Conference Series, Vol. 534, Protostars and Planets VII, ed. S.~{Inutsuka}, Y.~{Aikawa}, T.~{Muto}, K.~{Tomida}, \& M.~{Tamura}, 863, \dodoi{10.48550/arXiv.2203.10076}

\bibitem[{{Winn} \& {Fabrycky}(2015{\natexlab{a}})}]{WF15}
{Winn}, J.~N., \& {Fabrycky}, D.~C. 2015{\natexlab{a}}, \araa, 53, 409, \dodoi{10.1146/annurev-astro-082214-122246}

\bibitem[{{Winn} \& {Fabrycky}(2015{\natexlab{b}})}]{Winn15}
---. 2015{\natexlab{b}}, \araa, 53, 409, \dodoi{10.1146/annurev-astro-082214-122246}

\bibitem[{Zink {et~al.}(2021)Zink, Hardegree-Ullman, Christiansen, Bhure, Adkins, Petigura, Dressing, Crossfield, \& Schlieder}]{Zink21}
Zink, J.~K., Hardegree-Ullman, K.~K., Christiansen, J.~L., {et~al.} 2021, The Astronomical Journal, 162, 259, \dodoi{10.3847/1538-3881/ac2309}

\end{thebibliography}

\end{document}